\documentclass[%
 reprint,
superscriptaddress,
 amsmath,amssymb,comment,
 aps,
prl,
floatfix,
]{revtex4-2}

\usepackage{graphicx}% Include figure files
\usepackage{dcolumn}% Align table columns on decimal point
\usepackage{bm}% bold math
\usepackage{subfigure}
\usepackage{xcolor}
\usepackage{graphicx}
\usepackage[normalem]{ulem}
\usepackage{mathtools,amsmath,mathbbol}
\usepackage{amssymb}
\usepackage{float}
\usepackage{graphicx,siunitx}
\usepackage{subfigure,bm}
\usepackage[export]{adjustbox}
\usepackage{type1cm}
\usepackage{color,natbib}
\usepackage{empheq,wrapfig}
\usepackage{comment}

\usepackage{afterpage}

\usepackage{hyperref}
\hypersetup{
colorlinks = true,
urlcolor = magenta,
linkcolor = magenta,
citecolor = magenta
}

\begin{document}

\title{
Impact of granular inclusions on the phase behavior of colloidal gels
}

\author{Yankai Li}
\affiliation{%
 School of Engineering, The University of Edinburgh, King's Buildings, Edinburgh EH9 3FG, United Kingdom}%
\author{John R. Royer}%~\orcidlink{0000-0002-8368-7252}}
\affiliation{%
 SUPA, School of Physics and Astronomy, The University of Edinburgh, King's Buildings, Edinburgh EH9 3FD, United Kingdom}%
\author{Jin Sun}%~\orcidlink{0000-0001-6105-6743}}
\affiliation{%
 School of Engineering, The University of Edinburgh, King's Buildings, Edinburgh EH9 3FG, United Kingdom}%
 \author{Christopher Ness}%~\orcidlink{0000-0002-0842-2537}}
\affiliation{%
 School of Engineering, The University of Edinburgh, King's Buildings, Edinburgh EH9 3FG, United Kingdom}%

\date{\today}% It is always \today, today,

\begin{abstract}
Colloidal gels formed from small attractive particles are commonly used in formulations to keep larger components in suspension. However, despite extensive work characterizing unfilled gels, little is known about how larger inclusions alter the phase behavior and microstructure of the colloidal system. Here we use numerical simulations to examine how larger `granular' particles can alter the gel transition phase boundaries. We find two distinct regimes depending on both the filler size and native gel structure: a `passive' regime where the filler fits into already-present voids, giving little change in the transition, and an `active' regime where the filler no longer fits in these voids and instead perturbs the native structure. In this second regime the phase boundary is controlled by an effective colloidal volume fraction given by the available free volume.
%\footnote{{\bf Significance statement:Composites formed by mixing larger `granular' particles with attractive gel-forming colloids are ubiquitous in industry, and particularly relevant for battery manufacturing, but poorly understood. We use simulations to elucidate how these granular inclusions alter the transition between fluid and solid-like gel states. Though the interplay between the large grains and heterogeneous colloidal structures is complex, we discover a simple framework to describe the phase behavior based on the native gel structure, suggesting strategies to tune and optimise the properties of these granular-gel composites.}
%}
\end{abstract}

\maketitle

\section{Introduction}

Dispersions of attractive colloids can form solid-like gels characterized by a system-spanning network of arrested particles \cite{Trappe_gelrev_2004,emmag,Royall_Realspace_2021}. These colloidal gels are ubiquitous, encountered in disparate industries ranging from food and personal care products to building materials and catholyte slurries \cite{Cho_LiSlurry_2013,wei_biphasic_2015,Hawley_LiElec_2019,morelly}.
While there has been considerable progress in understanding the formation, structure and rheology of `model' colloidal gels formed from (nearly) uniformly sized spheres~\cite{Trappe_jamming_2001,Hsiao_iso_2012,Wang_adh_2019}, most practical gels are more complex. In particular, colloidal gels frequently serve as a carrier for larger, non-Brownian `granular' components (typical size $\gtrsim \SI{10}{\micro\meter}$). In such composites, the gel often acts as a rheology modifier to prevent sedimentation. In some applications the gel network itself may be desired, for example catholyte slurries for battery manufacturing rely on a conductive carbon black gel to provide connectivity between the active Li-ion `grains'. 

It is thus critical to understand how granular inclusions alter the colloidal gel phase. Recent work examining the influence of inclusions on gel rheology suggests they have a significant impact~\cite{claudia,st2}, even introducing new phenomena such as rheological bi-stability in these filled systems~\cite{jiang2022flow}. This previous work has largely focused on systems deep into the gel state,  so it remains unclear how granular inclusions alter the gel transition and phase behavior. 

For uniformly-sized colloidal spheres, the gelation phase boundaries depend on the colloid concentration and attraction strength, %For a given form of the interaction potential, this boundary can thus be defined by the colloid volume fraction, $\phi$, and the depth of the attractive well relative to the thermal energy,  $\epsilon/k_B T$. 
and there has been extensive work mapping these boundaries in a variety of systems~\cite{Grant_silicagel_1993,Verduin_adhs_1995,Poon_cpmix_1995,Serge_2001,Shah_depgel_2003,Sedgwick_lyso_2005,Lu_2008,Eberle_trgel_2011,Helgeson_nanoemu_2014,Whitaker_units_2019}. In depletion gels there is good agreement between gelation and the gas-liquid spinodal boundary~\cite{Tuinier_book_2011}, though there remains some debate concerning the generality of this agreement and the relative roles of percolation and clustering at the gel transition~\cite{Eberle_trgel_2011,Helgeson_nanoemu_2014,Whitaker_units_2019,Zhange_rigidperc_2019}. 

The inclusion of larger grains introduces additional control parameters which can potentially alter these phase boundaries. For simple `hard' grains interacting solely through their excluded volume, their influence will be set by the filler concentration and the size ratio $r_L/r_S$ between the large ($L$) grains and the smaller ($S$) colloids. These granular inclusions reduce the free volume available to the colloids relative to the total volume, but the interplay between the inclusions and gel structure is non-trivial, as colloidal gels can be heterogeneous on length scales $\gg r_S$ \cite{Fierro_2008, Zaccone_elast_2009,Royall_Realspace_2021}.

Here we characterise the influence of hard granular inclusions on the colloidal gel phase boundaries using numerical simulations, where the particle sizes, interactions and volume fractions can all be precisely varied. We find that the relative sizes of the inclusions and the void spaces present in the unfilled gels is the key parameter governing the phase behavior of the filled systems.

\section{Results and Discussion}

\begin{figure}
  \centering
  \includegraphics[width=0.48\textwidth]{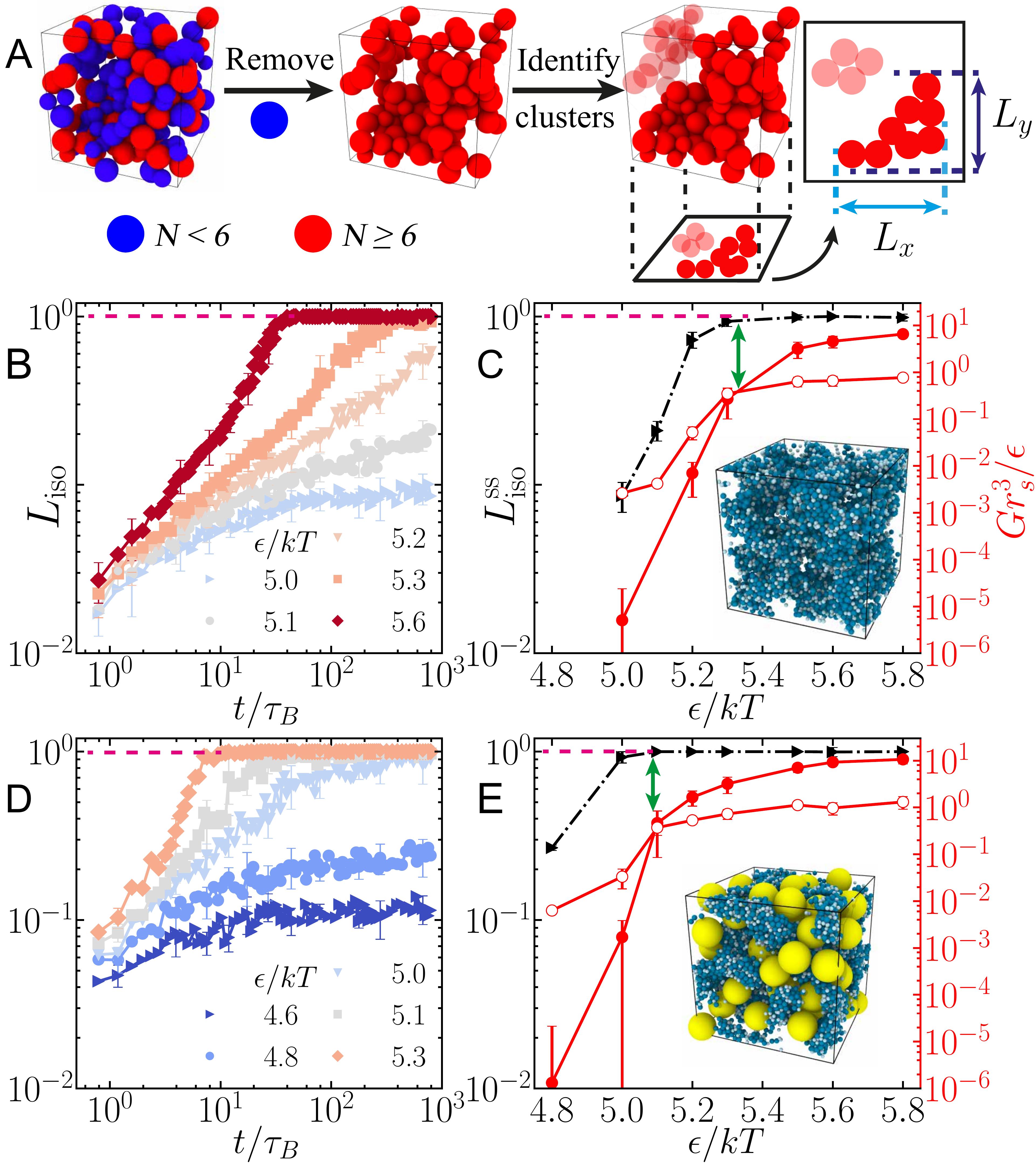}\\
  \caption{
  Isostaticity percolation and its relation to the mechanical response.
  (A) Procedure for determining $L_{\rm iso}$. Left to right: Isostatic colloids with $N\geq6$ contacts are isolated and contacting clusters identified. The isostatic length is computed from the span of the largest cluster, $L_{\rm iso} \equiv (L_x + L_y + L_z)/3L$.
  (B) Time evolution of $L_\text{iso}$ for an unfilled system ($\phi_S=0.2$, $\phi_L=0$) at varying attraction $\epsilon/kT$.
  (C) Steady-state $L^\text{ss}_\text{iso}$ (black, left axis) and viscoelastic moduli $G'$ and $G''$ (red filled and  open symbols, respectively, right axis) verses $\epsilon/kT$. Green arrow highlights the coincidence between the crossover $G'=G''$ and $L_\text{iso}^\text{ss}\simeq1$.
  Inset: rendering of a steady-state ($t/\tau_B \gg1)$ gel state for $\epsilon/kT=5.6$ with $\phi_S=0.2$, $\phi_L=0$.
%   Stress amplitude $\sigma$ at $5 kT$ (blue) and $6 kT$ (red), with frequency normalised by inertial time $\tau_i=m/6\pi\eta a$.
%   A low $\omega$ is chosen to guarantee that $\sigma$ is located in the linear (plateau) region at low (high) $\epsilon/kT$.
  (D) and (E) reproduce (B) and (C), respectively, for a filled system with $\phi_S=0.20$, $\phi_L=0.3$. Inset in (E) shows a rendering with granular inclusions (yellow) at $\epsilon/kT=5.6$. 
  Error bars in (B-E) indicate the standard deviation from 6 realisations.
  }\label{fig.fig1}
\end{figure}

We perform Langevin Dynamics using LAMMPS~\cite{Thompson_lammps_2022}, simulating the behavior of $10^4$ colloidal spheres (bidisperse with radii $r_S$ and $1.4r_S$) and a varying number of larger grains (monodisperse with radius $r_L$ varied between $8r_S-24r_S$). These simulations, detailed below in {\it Methods}, incorporate random Brownian forces, Stokes drag and inter-particle interactions, where we prescribe short range attraction between the small colloids and hard-sphere-like contact repulsion for interactions involving the larger grains. The system initially equilibrates without the colloidal attraction, then an interaction potential with depth $\epsilon/kT$ is turned on and the system evolved for up to $10^3 \tau_B$ to reach a steady state, with $\tau_B=6\pi\eta r_S^3/kT$ the Brownian diffusion time. To map the phase behavior of these systems, we vary $\epsilon/kT$ along with colloidal and granular volume fractions, $\phi_S = V_s/V$ and $\phi_L = V_L/V$. These volume fractions are defined by the volume occupied by the colloids (grains),$V_s$ ($V_L$), relative to the {\it total} volume of the cubic simulation box, $V=L^3$, so that increasing $\phi_L$ at fixed $\phi_S$ decreases the {\it free} volume available to the colloids.
 
\paragraph{Isostaticity percolation and mechanical response}
To explore how large granular inclusions alter the colloidal microstructure, we characterize networks of isostatic particles, defined as colloids with $\geq6$ contacts. This follows from the Maxwell criteria for stability in a system with pairwise central forces, requiring that constraints balance the degrees of freedom~\cite{Maxwell_1864}.
While this isostaticity criterion is typically considered globally in the context of granular packings~\cite{van_Hecke_2009}, it has been suggested that networks of locally isostatic particles control the rigidity of colloidal gels~\cite{Hsiao_iso_2012,Wang_2019}. Specifically, recent experimental work suggests the gel transition coincides with the formation of a percolating network of isostatic particles~\cite{Tsurusawa_iso_2019}.

We define contacts among attractive colloids as pairs (radii $r_i$ and $r_j$) within a separation $0.03(r_i + r_j)$ (see {\it Methods}).
To characterise the distance from isostaticity percolation, we first remove colloids with $<6$ contacts and then identify clusters of isostatic particles. The isostatic length, $L_\text{iso}$, gives the mean length of the largest isostatic cluster in all three spatial dimensions relative to the box size (Fig.~\ref{fig.fig1}A), so that $L_\text{iso}=1$ corresponds to isotropic isostaticity percolation. 

In pure colloidal suspensions ($\phi_L=0$), the colloids are initially well-dispersed  with $L_\text{iso}\approx 0$. Turning on the attraction $\epsilon$ at time $t=0$ causes clusters to form and grow, reflected in an increase in $L_\text{iso}(t)$ with time (Fig.~\ref{fig.fig1}B). As time progresses this initial growth slows and the isostatic length approaches a  plateau at some steady-state value $L^\text{ss}_\text{iso}$ at long times ($t\gtrsim 10^3 \tau_B$). The growth of $L_\text{iso}(t)$ depends on the attraction strength, with strongly attractive colloids rapidly reaching isostaticity percolation at $L^\text{ss}_\text{iso}=1$ while with weaker attraction $L_\text{iso}$ instead appears to plateau at some steady-state value $L^\text{ss}_\text{iso}<1$, short of the percolation threshold.

We apply small amplitude oscillatory shear $\gamma(t)=\gamma_0\sin\omega t$ to extract the viscoelastic moduli $G'$ and $G''$ for these steady-state structures (see Methods). We find a transition from liquid-like states ($G'<G''$) to solid-like states ($G'>G''$) as the attraction strength $\epsilon$ is increased (Fig.~\ref{fig.fig1}C), indicating the emergence of a solid-like gel. The crossover point where $G'=G''$ occurs as $L_\text{iso}^\text{ss}$ approaches unity at interaction energy $\epsilon^*$, indicating that the gel transition coincides with isostaticity percolation in agreement with Ref~\cite{Tsurusawa_iso_2019}. We verified this agreement holds for $\phi_S\leq0.4$.
% Under these conditions, the shear stress $\sigma$ in liquid-like states behaves as $\sigma\approx \omega\tau_i$
% and for solids as $\sigma=\text{constant}$.
%

%We verified that this result holds for $\phi_S\leq0.4$.

\paragraph{Role of granular inclusions}
The addition of granular inclusions gives qualitatively similar behavior in both the evolution of $L_\text{iso}$ with time and the mechanical response of the steady state structures, Figs.~\ref{fig.fig1}D,~E.  We again find good agreement between isostaticity percolation ($L_\text{iso}^\text{ss}\simeq 1$) and the emergence of mechanical rigidity ($G'=G''$), indicated by green arrows in Figs.~\ref{fig.fig1}C and E. Comparing these transition points for filled ($r_L = 8r_S$, $\phi_L=0.3$) and unfilled systems both at $\phi_S=0.2$, we find that the inclusions aid gelation with a reduced $\epsilon^*$ in the filled system. 

% Having established $L_\text{iso}^\text{ss}$ as our indicator of percolation ($L_\text{iso}^\text{ss}\simeq 1$) and mechanical rigidity,
% we next consider the role of large particles.
% It is established that for colloidal systems at a given $\epsilon/kT$,
% increasing $\phi_S$ takes the system across a `phase' boundary from liquid state to percolated structure~\cite{poon,phc1,phc2}.
% In our binary system,
% the transition can also (in some cases) be brought about by increasing $\phi_L$ (at fixed $\phi_S$, $\epsilon/kT$, Fig.~\ref{fig.fig2}a, b).

\begin{figure}
  \centering
  \includegraphics[width=0.48\textwidth]{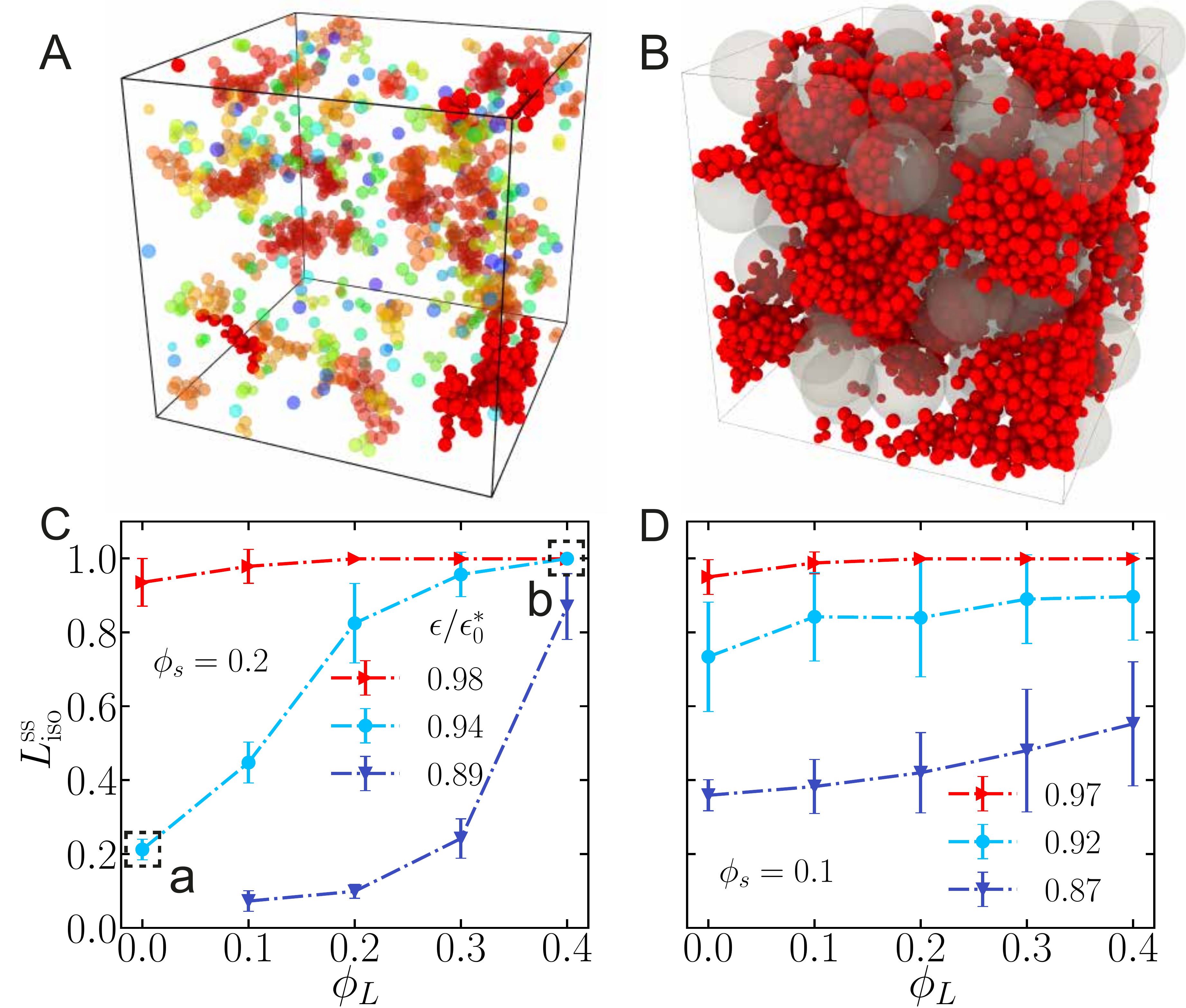}\\
  \caption{
  Contrasting effects of granular inclusions.
  Snapshots of steady-state ($t=10^3\tau_B$) configurations, with $\phi_S=0.2$ and $\epsilon=0.94\epsilon^*_0$ for
  (A) $\phi_L=0$ and; (B) $\phi_L=0.4$. Only isostatic colloids are rendered, with the largest cluster shown in red. Larger grains also shown (grey) in (B).
(C,D); 
$L_\text{iso}^\text{ss}(\phi_L)$ at varying $\epsilon$ for (C) $\phi_S=0.2$  and (D) $\phi_S=0.1$.
Values for $\epsilon$ given relative to the gelation point in the unfilled system, $\epsilon_0^*$, with $\epsilon_0^*=5.4kT$ for $\phi_S = 0.2$ and $\epsilon_0^*=6.3kT$ for $\phi_S = 0.1$.
Dashed squares a) and b) indicate states rendered in (A) and (B), respectively.
}
  \label{fig.fig2}
\end{figure}

%Renderings highlighting the isostatic colloids in these filled and un-filled systems (Fig.~\ref{fig.fig2}A,B) clarify this shift. 
To understand this shift in the gelation point, we examine how granular inclusions alter the structure and distribution of the isostatic colloidal particles. In a pure colloidal system below the gel transition, $\epsilon = \text{5.1} kT = 0.94 \epsilon^*$ and $\phi_S=0.2$, there are numerous disjoint clusters of isostatic particles (Fig.~\ref{fig.fig2}A) and the system remains well below the isostatic percolation threshold. The inclusion of the larger grains increases the number of isostatic colloidal particles, enabling them to instead form a large connected network which percolates across the sample for $\phi_L=0.4$ (Fig.~\ref{fig.fig2}B). 

For fixed $\phi_S=0.2$, we find that $L_\text{iso}^\text{ss}$ increases dramatically with increasing $\phi_L$ (Fig.~\ref{fig.fig2}C). Defining $\epsilon^*_0\equiv\epsilon^*(\phi_L=0)$ the critical interaction energy in the unfilled system, granular inclusions can take a system initially quite far from the gelation point ($\epsilon=0.89\epsilon^*_0$ and $L_\text{iso}^\text{ss}\approx 0$) nearly up to the transition point $L_\text{iso}^\text{ss}\lesssim 1$ as $\phi_L$ increases up to $\phi_L=0.4$. Increasing the interaction energy $\epsilon/\epsilon^*_0\to1$, the volume of granular filler needed to drive isostaticity percolation decreases, so that the gelation boundary $\epsilon^*$ where $L_\text{iso}^\text{ss}\to 1$ continuously shifts to lower values with increasing $\phi_L$.

Since the free volume available to the colloids decreases with increasing $\phi_L$, one might expect this increase in the number of isostatic particles, and hence an increase in $L_\text{iso}^\text{ss}$ in the filled system.   Furthermore, the gel phase boundary for unfilled systems, $\epsilon^*_0(\phi_S)$, is a decreasing function of $\phi_S$, at least at low to moderate concentrations~\cite{Grant_silicagel_1993,Verduin_adhs_1995,Poon_cpmix_1995,Serge_2001,Shah_depgel_2003,Sedgwick_lyso_2005,Lu_2008,Eberle_trgel_2011,Helgeson_nanoemu_2014,Whitaker_units_2019}. This suggests the possibility of capturing the shifting gel point with granular inclusions simply in terms of the reduced free volume available to the colloids. In this picture, one would expect the filler to have more pronounced effect at lower colloid concentrations, where the curve $\epsilon^*_0(\phi_S)$ is steepest. However, simulations with $\phi_S=0.1$
instead show the opposite, with the inclusion of large grains giving only a modest increase in $L_\text{iso}^\text{ss}$ (Fig.~\ref{fig.fig2}D) and minor shifts in the gelation phase boundary.

%$\epsilon_0$ for $\phi_S=0.2$ is 5.4 $kT$, for $\phi_S=0.1$ is 6.3 $kT$. A close $\epsilon/\epsilon_0$ value can not produce a similar curve of $L_\text{ios}^\text{ss}$ over $\phi_L$ at different $\phi_S$.

\begin{figure}[b]
  \centering
  \includegraphics[width=0.48\textwidth]{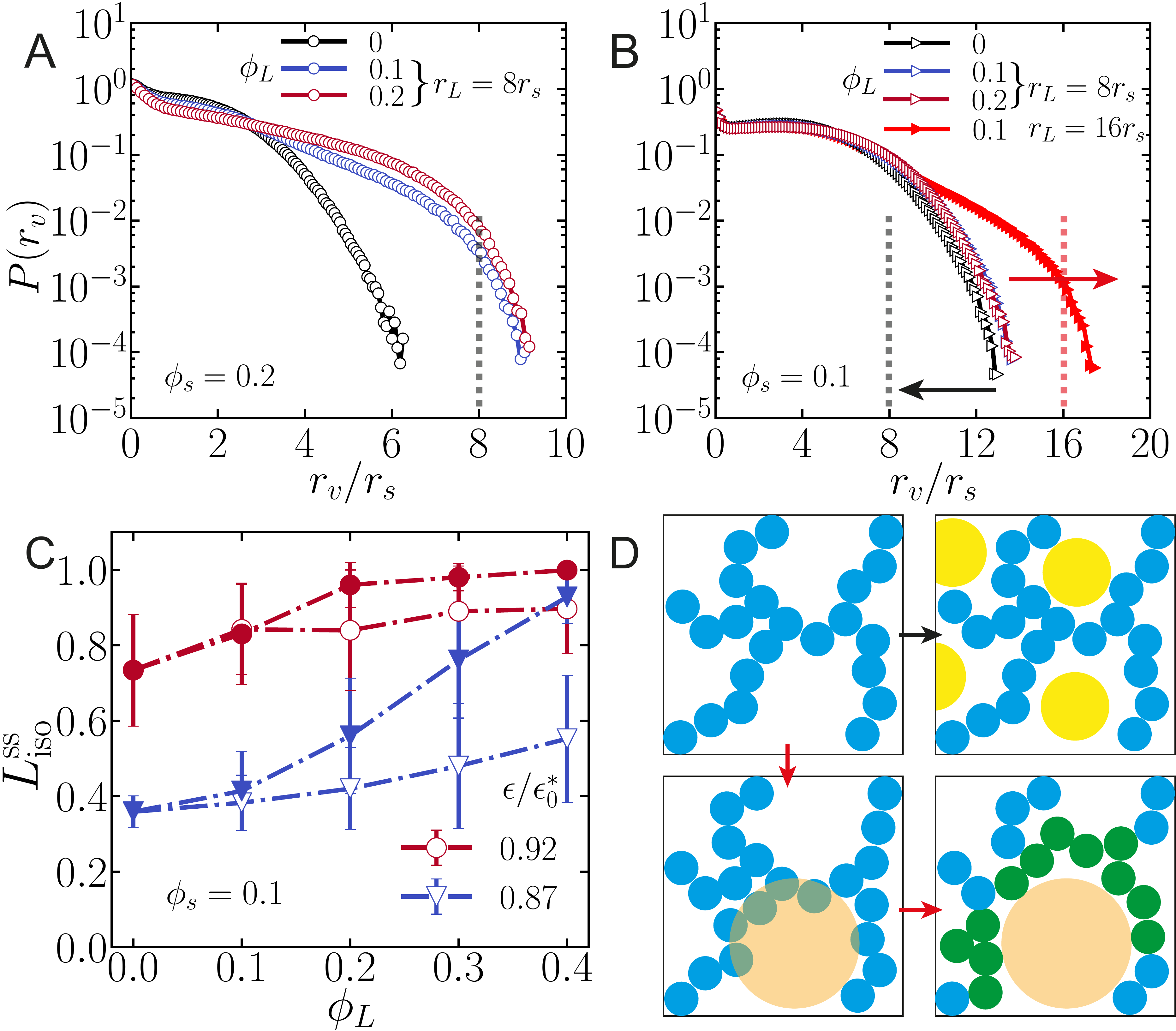}\\
  \caption{
Void-volumes in filled and unfilled systems near the gel boundary.
Normalized colloidal void size distribution $P(r_v)$ for (A) $\phi_S=0.2$ ($\epsilon= 5.3 kT \approx 0.98\epsilon_0$) and
(B) $\phi_S=0.1$ ($\epsilon = 6.3 kT \approx \epsilon_0^*$) at varying $\phi_L$.
The large particle size is $r_L=8r_S$ (highlighted by black dashed vertical lines), with the exception of the red filled symbols in (B) where $r_L =16r_S$ (highlighted by the red dashed vertical line).
(C) $L_\text{iso}^\text{ss}(\phi_L)$ for $\epsilon=5.8 kT$ (red circles) and $5.5 kT$ (blue triangles), with open symbols for $r_L=8r_S$ and filled symbols for $r_L=16r_S$.
Schematic cartoon in (D) presents a simplified picture to understand the influence (or lack thereof) of the granular filler on the gel structure, with large grains either fitting into `natural' voids in the gel, leaving the colloidal microstructure unaffected (black arrow), or forcing larger voids, which in turn distorts and compresses the colloidal phase (red arrows).
  }\label{fig.vvs}
\end{figure}

\paragraph{Fillers and voids}\label{vv}
To understand the reduced filler impact at low $\phi_S$, we look at how the microstructure of the colloidal phase varies with concentration. Specifically, we focus on the size distribution of the empty voids between the colloids at (or close to) the gel transition \cite{nick}. We compute this distribution by dividing the simulation volume into cubic cells (length $r_S$) and then finding the largest possible sphere (radius $r_v$) centered in each cell that avoids intersecting a {\it colloidal} particle (so that larger grains are treated as empty voids). Normalised histograms of these local void radii give the void size distribution $P(r_v)$. 

For an unfilled colloidal system with $\phi_S = 0.2$ close to the gel transition ($\epsilon = 5.3kT \approx 0.98 \epsilon^*_0$), this distribution is nearly flat up to $r_v\approx 3r_S$ and then falls off rapidly as $r_v$ increases further (Fig.~\ref{fig.vvs}A) with voids larger than $r_v\approx 6r_S$ exceedingly rare. While precisely characterizing the rare-event tails in $P(r_v)$ would require significant computational effort, we can define an effective maximum void size $P(r_v^\text{max}) = 10^{-4}$, as voids larger then this are effectively absent in our observed configurations.  Adding larger granular inclusions, with $r_L = 8r_S > r_v^\text{max}$, perturbs the colloidal microstructure and shifts the shoulder in $P(r_v)$ to higher radii $\approx r_L$, reflecting the voids created by the large grains.

\begin{figure}
  \centering
  \includegraphics[width=0.4\textwidth]{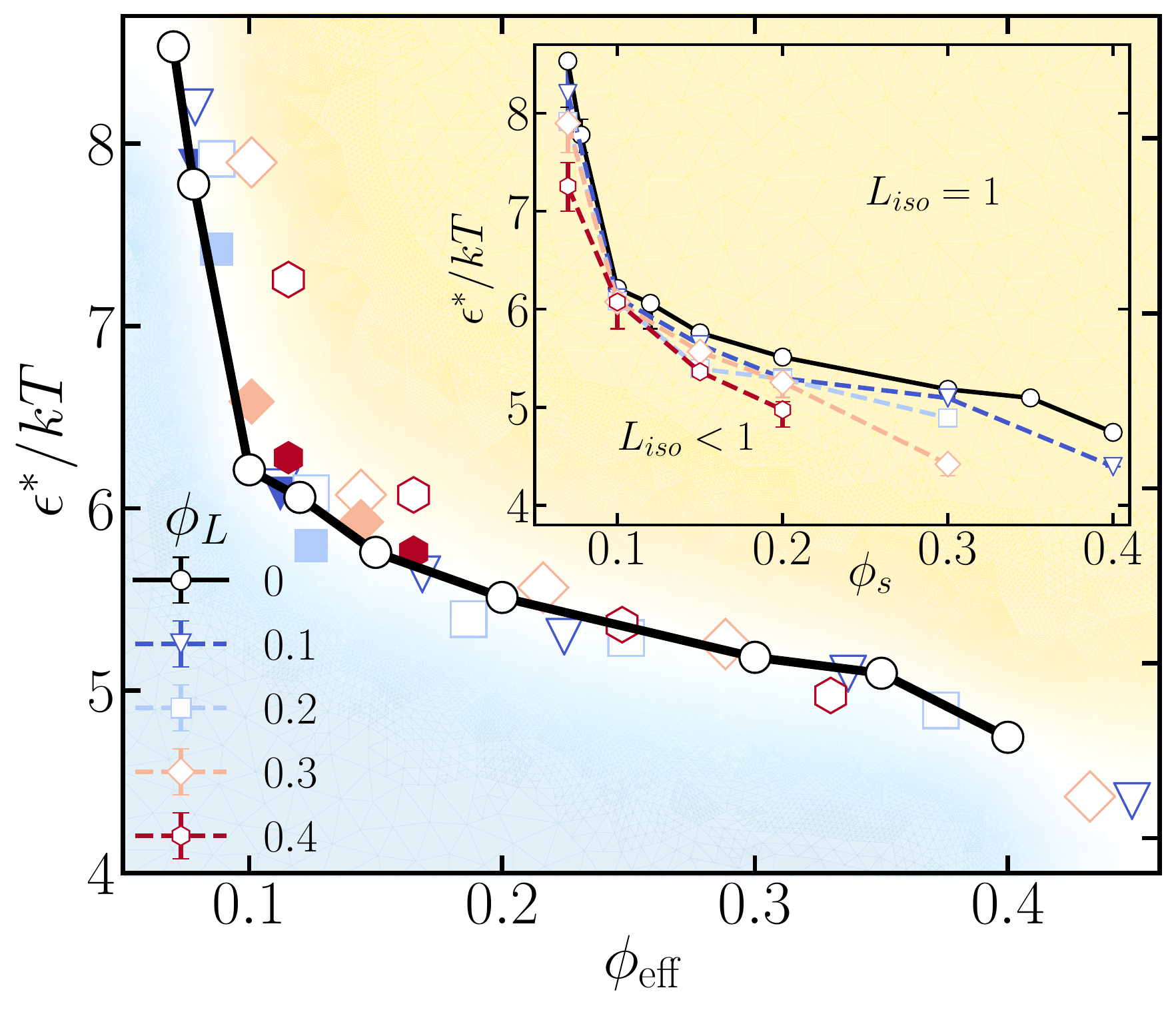}\\
  \caption{Isostatic percolation boundaries $\epsilon^*$ varying $\phi_S$ and $\phi_L$. Inset: Critical attractive energy $\epsilon^*(\phi_S)$ for varying $\phi_L$ (indicated in main panel legend) for $r_L = 8r_S$. Main panel: Open symbols show same data as inset, now plotted against $\phi_\text{eff}=\phi_S/(1-\phi_L)$. Solid symbols show results for larger grains and smaller $\phi_S$, with $r_L = 16r_S$ for $\phi_S = 0.1$ and $r_L = 24r_S$ for $\phi_S = 0.07$ so that $r_L > r_v^\text{max}$ in both cases.
  Error bars on $\epsilon^*$ computed using error bars on $L_\text{iso}^\text{ss}$ to obtain upper and lower bounds for the point the isostatic length exceeds our threshold $L_\text{iso}^\text{ss}=L_\text{iso}^*=0.99$ and are shown only in the inset for clarity.
  }\label{fig.bd}
\end{figure}

However, reducing the colloid concentration to $\phi_S=0.1$ gives a significantly wider distribution of void sizes in the unfilled system, with the shoulder in $P(r_v)$ now around $r_v=8r_S$ and $r_v^\text{max} \approx 12r_S$ (Fig.~\ref{fig.vvs}B). This indicates there are `pre-existing' voids which can accommodate the larger $r_L=8r_S$ grains without forcing a significant change in the colloidal microstructure. Indeed, in contrast to the large shift seen for $\phi_S=0.2$, here increasing $\phi_L$ has only a minor impact on the shape of $P(r_v)$. 
This suggests a picture where dilute gels with $r_v > r_L$ can effectively form around the immobile granular inclusions, forming a network of colloid-colloid contacts that is essentially indistinguishable from the unfilled case. With no change in this network, $L_\text{iso}$ is unaffected by the granular inclusions and there is little shift in the gel phase boundary. We would thus expect that increasing $r_L$ above the characteristic void size would result in a more pronounced filler-effect on $L_\text{iso}^\text{ss}$ and the gelation point in these more dilute gels.

We find this is indeed the case when $r_L$ is increases from $8r_S$ to $16r_S$. These larger grains now notably shift $P(r_v)$ to higher values (compare open and filled symbols in Fig.~\ref{fig.vvs}B), and also now find a clear increase in $L_\text{iso}^\text{ss}(\phi_L)$ (Fig.~\ref{fig.vvs}C), similar to the impact of the $r_L=8r_S$ grains in the $\phi_S=0.2$ system with smaller voids. This supports a simplified picture of the interplay between the granular inclusions and attractive colloids (Fig.~\ref{fig.vvs}D); grains with $r_L<r_v$ have little impact on the gel structure or transition point as they easily fit into the gel voids, while grains with $r_L>r_v$ distort and compress colloidal structures, increasing $L_\text{iso}^\text{ss}$ and reducing the attraction needed to form the gel.

\paragraph{Phase Diagram}

Having detailed the influence of granular inclusions at two specific $\phi_S$, we now map out the gel phase boundaries over a range of $\phi_S$ and $\phi_L$. Having demonstrated good agreement between the rigidity onset and isostaticity percolation, we use $L_\text{iso}^\text{ss}$ to classify states. Carrying out a series of simulations at varying $\epsilon$ for given $\phi_S$ and $\phi_L$, we define the transition point $\epsilon^*$ as the point $L_\text{iso}^\text{ss}=L_\text{iso}^*=0.99$ using linear interpolation between the finite $\epsilon$ steps. The need for a threshold $L_\text{iso}^*\lesssim1$ reflects ambiguities in counting bonds which cross our periodic boundaries. We verified our results are not sensitive to the precise threshold, with values $0.99\leq L_\text{iso}^*\leq0.999$ giving nearly indistinguishable phase boundaries.

For the range of $\phi_S$ explored here (up to $\phi_S=0.4$), we find the isostatic percolation boundary $\epsilon^*(\phi_S)$ monotonically decreases with increasing $\phi_S$ (Fig.~\ref{fig.bd} inset). For fixed $r_L = 8r_S$, increasing $\phi_L$ at fixed $\phi_S$ generally shifts this boundary to lower attraction strength.

Instead plotting these isostatic percolation boundaries as a function of an effective volume fraction $\phi_\text{eff} \equiv \phi_S/(1-\phi_L)$, giving the volume fraction of the small colloids relative to the free volume excluding the large grains $(1-\phi_L)V$, we find reasonable collapse for $\phi_\text{eff}\gtrsim 0.2$ (Fig.~\ref{fig.bd} main panel). This suggests that the filler-induced shifts in the phase boundary can be understood solely through the reduction in free volume available to small colloids, so that adding larger grains is effectively equivalent to shrinking the box volume. However, for $\phi_S \lesssim 0.1$ we find this collapse breaks down, with points at higher $\phi_L$ lying clearly above the $\phi_L=0$ boundary. This is consistent with the behavior seen in Fig.~\ref{fig.fig2}D, where the granular inclusions only have a minor effect on $L_\text{iso}$ for $\phi_S=0.1$ compared to the significant enhancement seen at a higher $\phi_S=0.2$. 

Increasing the size of the large particles to ensure $r_L > r_v^\text{max}$, in this case setting $r_L = 16r_S$ for $\phi_S = 0.1$ and $r_L = 24r_S$ for $\phi_S = 0.07$,  we find that this collapse can be recovered (filled symbols in Fig.~\ref{fig.bd}). We thus see that when the granular inclusions are significantly larger than the typical voids in the unfilled gel, so that they force a notable change in the gel microstructure, the influence of the voids can be captured by the effective free volume available to the small colloids. With smaller grains this effect is diminished, with the phase boundary instead largely independent of the filler concentration.

\section{Conclusions}

Using Langevin dynamics simulations, 
we have mapped out the influence of larger granular inclusions on isostaticity percolation and the gel transition in suspensions of smaller attractive colloids. Varying the volume fractions of both species,
we demonstrated two distinct regimes:
({\it i}) a `passive void-filling' regime, where the granular inclusions can fit into already-present voids within the gel, so that the microstructure is effectively unchanged and the gel transition governed almost solely by $\phi_S$ and
({\it ii}) an `active void-enhancing' regime where the granular inclusions perturb the gel structure by forcing larger voids and the gel transition governed by an effective volume fraction $\phi_\text{eff}={\phi_S}/(1-\phi_L)$.
These two limiting regimes are differentiated by the size ratio of the larger grains $r_L$ and shape of the gel void size distribution $P(r_v)$, so that anticipating the impact of the granular filler requires detailed characterization of the unfilled gel structure. 

There is relatively little experimental work examining the phase behavior of filled colloidal gels, though a recent study using a battery electrode slurry (a carbon black gel with $\approx \SI{10}{\micro\meter}$ granular inclusions) found little change with addition of the granular particles~\cite{morelly}. Given the very low gel point (occurring at $\phi_S\approx0.02$), it is plausible that the carbon black gel contains sufficiently large voids to place this system in regime ({\it i}), though detailed characterisation of the gel structure would be required to confirm this. Our results should be particularly relevant to battery slurry formulation and electrode fabrication, where particle connectivity is key for performance~\cite{Hawley_LiElec_2019}, providing a road map to match the native gel structure and filler properties to tune the electrode microstructure.  

%%\cite{claudia}, and contribute directly to understanding technological applications such as semi-solid flow cells, of which the electrode micro-structure determines ion and electron transport, as well as battery capacity \cite{btt}, but lacks a thorough investigation. 

% \appendix

\section{Methods}
\small{

We simulate the trajectories of $10^4$ colloidal and a smaller number of granular spheres in a periodic box (volume $V$) according to the Langevin equation, which for particle $i$ reads
%\begin{equation}
$
m_i d\textbf{U}_i/dt=\textbf{F}_i^H+\textbf{F}_i^B+\textbf{F}_i^P\text{,}
$
%\end{equation}
with $m_i$ and $\textbf{U}_i$ the particle mass and velocity respectively.
The hydrodynamic force 
%\begin{equation}\label{eq.hydro}
$
    \textbf{F}_i^H=-6\pi\eta r_i(\textbf{U}_i-\textbf{U}^\infty_i),
$
%\end{equation}
captures Stokes drag on a sphere with radius $r_i$, with $\textbf{U}^\infty_i$ the background fluid velocity (generally set to 0 except under oscillatory shear).

Brownian forces are generated as
%\begin{equation}\label{eq.brown}
$
\textbf{F}_i^B=\sqrt{12\pi\eta r_i kT/\Delta t}\textbf{R},
$
%\end{equation}
where $\Delta t$ is the timestep, $kT$ the thermal energy and the elements of the vector $\textbf{R}$ are drawn from a Gaussian distribution with zero-mean and no time correlation.
%$\overline{R(t)}=0,\ \overline{R(t)R(t^\prime)}=\delta(t-t^\prime)$.
The characteristic diffusive timescale for a particle with radius $r_i$ is thus $6\pi\eta r_i^3/kT$. To avoid crystallisation in the small colloids, we use a binary size mixture with radii $r_S$ and $1.4r_S$, while the larger grains are monodisperse with radius $r_L$ varying from $8r_S$ to $24r_S$. Since the diffusion time scales as $r_i^3$, for the larger grains it is $>500$ times longer than the colloidal timescale $\tau_B=6\pi\eta r_S ^3/kT$, so that even though Brownian forces are applied uniformly to all particles the larger grains are effectively non-Brownian.

Colloids (labeled $i$ and $j$) at a distance $r$ and surface-to-surface separation $\delta_{ij} = r-(r_i+r_j)$ interact via a Morse potential, giving a force
%\begin{equation}\label{eq.brown}
$
    \textbf{F}_{ij}^p=\epsilon \kappa_{ij}  e^{-\kappa_{ij}\delta_{ij}}(e^{-\kappa_{ij}\delta_{ij}}-1)\textbf{n}_{ij}\text{}
$
%\end{equation}
with $\textbf{n}_{ij}$ the center-to-center unit vector.
This potential gives finite-ranged attraction, and repulsion for overlapping particles ($\delta_{ij}<0$).
The interaction length scale is set as $\kappa_{ij}^{-1}=(r_i+r_j)/200$ to give short-ranged attraction.
We evaluate the force when $\delta_{ij}<0.03(r_i+r_j)$ (following conventional practice, see e.g.~\cite{sciortino2004equilibrium}), beyond which the attractive force is $<1\%$ of its maximum value.
We defined contacts using the same threshold of $\delta_{ij}$, having verified that a more stringent criteria does not affect any of the conclusions we draw.
The depth of the attractive potential $\epsilon$ is varied between $kT$ and $20kT$, with a variable step size to refine our estimates of the gelation point.
Colloid-granular and granular-granular contact forces are modeled as linear springs
%\begin{equation}\label{eq.hookean}
$
    \textbf{F}_{ij}^p=- k_n\delta_{ij}\textbf{n}_{ij},
$
%\end{equation}
with a stiffness $k_n$ set sufficiently large to approximate hard-sphere interactions (i.e. $k_nr_i^2\gg\epsilon$, $kT$). The timestep $\Delta t$ is set to $10^{-4}$, substantially smaller than $\tau_B$ and $\sqrt{m/k_n}$.

We characterise the mechanical response of steady-state structures by applying an oscillatory shear $\textbf{U}^\infty_i(t,y_i) = y_i\gamma_0\sin\omega t $ and turning off the Brownian forces. The bulk shear stress is computed as $\sigma_{ij} = V^{-1} \sum r_{ij} F^\text{tot}_{ij}$ (with the sum being over all interacting pairs), and the viscoelastic moduli $G'$ and $G''$ computed from the Fourier transform of $\sigma_{xy}$ averaged over 50 shear cycles.
All results for both the elastic moduli and $L_\text{iso}$ are averages from 6 independent realisations with randomized granular and colloidal initial positions.

}

\section{Acknowledgments}
YL is funded by the China Scholarship Council (CSC)
and by the University of Edinburgh through a School of Engineering studentship.
C.N. acknowledges support from the Royal Academy of Engineering under the Research Fellowship scheme.
The work was supported by the UK Engineering and Physical Sciences Research Council under grant EP/N025318/1.
For the purpose of open access, the authors have applied a Creative Commons Attribution (CC BY) licence to any Author Accepted Manuscript version arising from this submission.
YL, JR, JS and CN planned the research;
YL carried out the research;
YL, JR and CN wrote the manuscript.
The data used to generate the figures in this article is available via Edinburgh DataStore at XXX.

\bibliography{apssamp}% Produces the bibliography via BibTeX.

\end{document}